\documentstyle[twoside]{article}

\catcode`\@=11
\long\def\@makefntext#1{
\protect\noindent \hbox to 3.2pt {\hskip-.9pt  
$^{{\eightrm\@thefnmark}}$\hfil}#1\hfill}		

\def\thefootnote{\fnsymbol{footnote}}
\def\@makefnmark{\hbox to 0pt{$^{\@thefnmark}$\hss}}	
	
\def\ps@myheadings{\let\@mkboth\@gobbletwo
\def\@oddhead{\hbox{}
\rightmark\hfil\eightrm\thepage}   
\def\@oddfoot{}\def\@evenhead{\eightrm\thepage\hfil
\leftmark\hbox{}}\def\@evenfoot{}
\def\sectionmark##1{}\def\subsectionmark##1{}}

\oddsidemargin=\evensidemargin
\renewcommand{\thefootnote}{\fnsymbol{footnote}}

\newcounter{sectionc}\newcounter{subsectionc}\newcounter{subsubsectionc}
\renewcommand{\section}[1] {\vspace{12pt}\addtocounter{sectionc}{1} 
\setcounter{subsectionc}{0}\setcounter{subsubsectionc}{0}\noindent 
	{\tenbf\thesectionc. #1}\par\vspace{5pt}}
\renewcommand{\subsection}[1] {\vspace{12pt}\addtocounter{subsectionc}{1} 
	\setcounter{subsubsectionc}{0}\noindent 
	{\bf\thesectionc.\thesubsectionc. {\kern1pt \bfit #1}}\par\vspace{5pt}}
\renewcommand{\subsubsection}[1] {\vspace{12pt}\addtocounter{subsubsectionc}{1}
	\noindent{\tenrm\thesectionc.\thesubsectionc.\thesubsubsectionc.
	{\kern1pt \tenit #1}}\par\vspace{5pt}}
\newcommand{\nonumsection}[1] {\vspace{12pt}\noindent{\tenbf #1}
	\par\vspace{5pt}}

\newcounter{appendixc}
\newcounter{subappendixc}[appendixc]
\newcounter{subsubappendixc}[subappendixc]
\renewcommand{\thesubappendixc}{\Alph{appendixc}.\arabic{subappendixc}}
\renewcommand{\thesubsubappendixc}
	{\Alph{appendixc}.\arabic{subappendixc}.\arabic{subsubappendixc}}

\renewcommand{\appendix}[1] {\vspace{12pt}
        \refstepcounter{appendixc}
        \setcounter{figure}{0}
        \setcounter{table}{0}
        \setcounter{lemma}{0}
        \setcounter{theorem}{0}
        \setcounter{corollary}{0}
        \setcounter{definition}{0}
        \setcounter{equation}{0}
        \renewcommand{\thefigure}{\Alph{appendixc}.\arabic{figure}}
        \renewcommand{\thetable}{\Alph{appendixc}.\arabic{table}}
        \renewcommand{\theappendixc}{\Alph{appendixc}}
        \renewcommand{\thelemma}{\Alph{appendixc}.\arabic{lemma}}
        \renewcommand{\thetheorem}{\Alph{appendixc}.\arabic{theorem}}
        \renewcommand{\thedefinition}{\Alph{appendixc}.\arabic{definition}}
        \renewcommand{\thecorollary}{\Alph{appendixc}.\arabic{corollary}}
        \renewcommand{\theequation}{\Alph{appendixc}.\arabic{equation}}
        \noindent{\tenbf Appendix \theappendixc #1}\par\vspace{5pt}}
\newcommand{\subappendix}[1] {\vspace{12pt}
        \refstepcounter{subappendixc}
        \noindent{\bf Appendix \thesubappendixc. {\kern1pt \bfit #1}}
	\par\vspace{5pt}}
\newcommand{\subsubappendix}[1] {\vspace{12pt}
        \refstepcounter{subsubappendixc}
        \noindent{\rm Appendix \thesubsubappendixc. {\kern1pt \tenit #1}}
	\par\vspace{5pt}}

\newcommand{\textlineskip}{\baselineskip=13pt}
\newcommand{\smalllineskip}{\baselineskip=10pt}

\def\eightcirc{
\begin{picture}(0,0)
\put(4.4,1.8){\circle{6.5}}
\end{picture}}
\def\eightcopyright{\eightcirc\kern2.7pt\hbox{\eightrm c}} 

\newcommand{\copyrightheading}[1]
	{\vspace*{-2.5cm}\smalllineskip{\flushleft
	{\footnotesize International Journal of Modern Physics A, #1}\\
	{\footnotesize $\eightcopyright$\, World Scientific Publishing
	 Company}\\
	 }}

\def\abstracts#1#2#3{{
	\centering{\begin{minipage}{4.5in}\baselineskip=10pt\footnotesize
	\parindent=0pt #1\par 
	\parindent=15pt #2\par
	\parindent=15pt #3
	\end{minipage}}\par}} 

\renewenvironment{thebibliography}[1]
	{\frenchspacing
	 \ninerm\baselineskip=11pt
	 \begin{list}{\arabic{enumi}.}
	{\usecounter{enumi}\setlength{\parsep}{0pt}
	 \setlength{\leftmargin 12.7pt}{\rightmargin 0pt} 
	 \setlength{\itemsep}{0pt} \settowidth
	{\labelwidth}{#1.}\sloppy}}{\end{list}}

\newcounter{itemlistc}
\newcounter{romanlistc}
\newcounter{alphlistc}
\newcounter{arabiclistc}

\newcommand{\fcaption}[1]{
        \refstepcounter{figure}
        \setbox\@tempboxa = \hbox{\footnotesize Fig.~\thefigure. #1}
        \ifdim \wd\@tempboxa > 5in
           {\begin{center}
        \parbox{5in}{\footnotesize\smalllineskip Fig.~\thefigure. #1}
            \end{center}}
        \else
             {\begin{center}
             {\footnotesize Fig.~\thefigure. #1}
              \end{center}}
        \fi}

\newcommand{\tcaption}[1]{
        \refstepcounter{table}
        \setbox\@tempboxa = \hbox{\footnotesize Table~\thetable. #1}
        \ifdim \wd\@tempboxa > 5in
           {\begin{center}
        \parbox{5in}{\footnotesize\smalllineskip Table~\thetable. #1}
            \end{center}}
        \else
             {\begin{center}
             {\footnotesize Table~\thetable. #1}
              \end{center}}
        \fi}

\def\@citex[#1]#2{\if@filesw\immediate\write\@auxout
	{\string\citation{#2}}\fi
\def\@citea{}\@cite{\@for\@citeb:=#2\do
	{\@citea\def\@citea{,}\@ifundefined
	{b@\@citeb}{{\bf ?}\@warning
	{Citation `\@citeb' on page \thepage \space undefined}}
	{\csname b@\@citeb\endcsname}}}{#1}}

\newif\if@cghi
\def\cite{\@cghitrue\@ifnextchar [{\@tempswatrue
	\@citex}{\@tempswafalse\@citex[]}}
\def\citelow{\@cghifalse\@ifnextchar [{\@tempswatrue
	\@citex}{\@tempswafalse\@citex[]}}
\def\@cite#1#2{{$\null^{#1}$\if@tempswa\typeout
	{IJCGA warning: optional citation argument 
	ignored: `#2'} \fi}}

\def\pmb#1{\setbox0=\hbox{#1}
	\kern-.025em\copy0\kern-\wd0
	\kern.05em\copy0\kern-\wd0
	\kern-.025em\raise.0433em\box0}

\def\fnm#1{$^{\mbox{\scriptsize #1}}$}
\def\fnt#1#2{\footnotetext{\kern-.3em
	{$^{\mbox{\scriptsize #1}}$}{#2}}}

\def\fpage#1{\begingroup
\voffset=.3in
\thispagestyle{empty}\begin{table}[b]\centerline{\footnotesize #1}
	\end{table}\endgroup}

\def\runninghead#1#2{\pagestyle{myheadings}
\markboth{{\protect\footnotesize\it{\quad #1}}\hfill}
{\hfill{\protect\footnotesize\it{#2\quad}}}}
\font\tenrm=cmr10
\font\tenit=cmti10 
\font\tenbf=cmbx10
\font\bfit=cmbxti10 at 10pt
\font\ninerm=cmr9

\font\eightrm=cmr8

\def\qed{\hbox{${\vcenter{\vbox{			
   \hrule height 0.4pt\hbox{\vrule width 0.4pt height 6pt
   \kern5pt\vrule width 0.4pt}\hrule height 0.4pt}}}$}}

\renewcommand{\thefootnote}{\fnsymbol{footnote}}	

\input epsf.tex
\def\greatersim{\ \hbox{\raise 2pt \hbox{$>$} \kern -13pt
                     \lower 3pt \hbox{$\sim$}}\ }
\def\desepsf(#1 width #2){\epsfxsize=#2 \epsfbox{#1}}
\begin{document}

\runninghead{Radiative quarkonia decays 
$\ldots$} {Radiative quarkonia decays 
$\ldots$}

\normalsize\textlineskip
\setcounter{page}{1}

\copyrightheading{}			

\vspace*{0.88truein}

\fpage{1}
\centerline{\bf THE ENDPOINT REGION IN RADIATIVE QUARKONIA 
DECAYS\fnm{a,}\fnt{a}{Talk at the 
DPF2000 Meeting, Ohio State University, 9-12 August 
2000.}\fnm{b}\fnt{b}{Work funded in part by the US Department of Energy.}}
\vspace*{0.37truein}
\centerline{\footnotesize F. HAUTMANN}
\vspace*{0.035truein}
\centerline{\footnotesize\it  Department of Physics, Pennsylvania 
State University, University Park PA 16802}
\vspace*{10pt}

\vspace*{0.21truein}
\abstracts{
We consider  the inclusive  radiative decays $quarkonium$ $\to$ $\gamma$ 
$+$ $hadrons$ and examine the effects of  soft QCD radiation on the 
photon energy spectrum near the endpoint. 
}{}{}

\textlineskip			
\vspace*{12pt}			


\textheight=7.8truein
\setcounter{footnote}{0}
\renewcommand{\thefootnote}{\alph{footnote}}

\vspace*{12pt}

The use of perturbation theory  
to describe radiative decays of 
quarkonia  is based on the fact that, as long 
as the velocity $v$ of the heavy quark  is small, these decay processes 
involve two widely separated distance scales: the  
scale $1/ (m v^2)$ over which the quark and antiquark bind into the  
quarkonium and the 
scale $1/m$ over which the quark-antiquark pair decays 
(with $m$ the heavy quark mass). 
By expanding about the nonrelativistic limit $v \to 0$, one may treat 
the process as  the  product of a long-distance, 
nonperturbative factor, containing all of 
the bound-state dynamics, and a short-distance factor, 
 describing the annihilation of the  
heavy quark pair and  computable as a 
power series expansion in $\alpha_s$. 

However, near the exclusive boundary of the phase space, 
 where the photon's energy $E_\gamma$  
 approaches its kinematic limit, 
both the expansion to fixed 
order in $\alpha_s$ and the expansion to fixed order in $v$ 
become inadequate  to represent correctly the physics of the decay. 
On  one  hand, potentially large terms in $\ln (1-z)$ (with  
 $z=E_\gamma/m$)    
appear in the coefficients   of  
the expansion in $\alpha_s$ to all orders\cite{Photia}.  
On the other  hand,  classes of relativistic corrections  
 that  by power counting are higher order 
in $v$   become enhanced by powers of $1 / 
\alpha_s$\cite{rothwise,malpet,wolf}. 
These behaviors depend on   
the soft color radiation associated with the decay.   
In both cases, 
reliable results can only be obtained after  resummation of the enhanced 
contributions.  
An analogous sensitivity  
to  infrared dynamics  is  seen  
  in  calculations that 
incorporate  models for  
the hadronization of partons\cite{montfield,consfie,Field}:      
nonperturbative contributions are found to be  essential to 
describe the photon energy spectrum\cite{earlier,cleo} near the endpoint.  

In this talk we 
outline  a treatment of the infrared  radiation near the kinematic 
boundary, including soft-gluon coherence effects, and  
give an argument for the cancellation of all the 
corrections  in $\ln (1-z)$ in the short distance coefficient for  
the color-singlet Fock state in the quarkonium\cite{egmond}.  
 We focus on  processes in which the photon is directly 
coupled to the heavy quarks, because 
near the endpoint these contributions dominate the 
contributions in which 
the photon is produced by fragmentation\cite{mont}  
of light quarks and gluons. 

Consider the lowest order contribution for 
a $^3 S_1$ quarkonium 
$H$ of mass $M$ in a color singlet state, 
 $H \to \gamma g g$. 
 The  normalized photon spectrum at this order is\cite{Brod}  
\begin{equation}
\label{borng}
{ 1 \over \Gamma_0} \ {{d \Gamma_0} \over { d z} } = 
{1 \over {\pi^2 -9}} \ \int_0^1 d x_1 
  \int_0^1 d x_2   \ {\cal M}_0 (x_1 , x_2, z) \ \delta(z - 2 + x_1 + x_2)
\  \Theta(x_1+x_2 -1) 
\hspace*{0.1 cm} , 
\end{equation}
with 
\begin{equation}
\label{tree}
{\cal M}_0 (x_1 , x_2, z) = 
  {{(1-x_1)^2} \over { z^2 x_2^2}} + 
{{(1-x_2)^2} \over { z^2 x_1^2}} + {{(1-z)^2} \over { x_1^2 x_2^2}} 
\hspace*{0.4 cm} .  
\end{equation}
Here $x_1, x_2$ denote  gluon energy fractions, $x_i=2 E_i / M$. 
The result (\ref{borng}) 
is well approximated by a spectrum rising linearly with $z$. 
It goes to a constant as $z \to 1$.

To higher perturbative orders, 
 logarithmic  corrections  arise 
from soft and collinear gluon radiation. By power counting 
 the  leading behavior of  the spectrum  
as $z \to 1$ is  of the type\cite{Photia}  
\begin{equation}
\label{hologs}
{1 \over \Gamma} \ 
{{d \Gamma} \over { d z} } \sim 
{\mbox{const.}} \ + \sum_{k=1}^{\infty} 
c_k \ \alpha_s^k \ \ln^{2 k} (1-z) \;\; ,  \;\;\;\;\; z \to 1 
\hspace*{0.4 cm} . 
\end{equation}
The coherent branching algorithm\cite{coh} provides a method to 
 evaluate the relevant 
multiparton  matrix elements in the 
soft and collinear regions.  
 This algorithm effectively replaces 
the calculation of higher-loop Feynman graphs by the calculation 
of tree-level 
graphs, in which  the angular phase space 
 is subject to  ordering constraints at each branching. The basis 
for this method is the coherence property of soft gluon 
emission. 

This method has been developed and applied 
to the study of event-shape observables in $e^+ e^-$ annihilation\cite{cttw}. 
The photon spectrum in decays of quarkonia can be 
treated in a similar manner. Following this approach, 
  the photon spectrum can be 
expressed, to all orders in $\alpha_s$ and to 
the leading and next-to-leading  accuracy in $\ln (1-z)$, 
in terms of the on-shell amplitude for the tree-level process  
$H(P) \to \gamma(k) g(k_1) g(k_2)$   
and the mass distributions $J_g$ for the 
time-like 
jets  defined by the  branching algorithm\cite{egmond}: 
\begin{eqnarray}
\label{branchfact}
{1 \over \Gamma} \ 
{{d \Gamma} \over { d z} } &=& 
\ \int  {{ d^4 k} \over {(2 \pi)^3}} 
{{ d^4 k_1} \over {(2 \pi)^3}} {{ d^4 k_2} \over {(2 \pi)^3}} 
\nonumber\\
&\times& 
(2 \pi)^4 \delta^4 (P-k-k_1-k_2) 
\delta \left(z - {{2 P \cdot k} \over M^2} \right)
\delta_+ (k^2) 
\nonumber\\
&\times& 
  {\cal M}^{\rm{(tree)}} (P, k_1 , k_2) \ 
  J_g \left( (k_1+k_2)^2 , k_1^2 \right) \ 
  J_g \left( (k_1+k_2)^2 , k_2^2 \right) 
\hspace*{0.6 cm} .   
\end{eqnarray}

The precise definition of 
$J (p^2,k^2)$ is given in ref.\cite{cttw}. The first argument in 
$J$ is the coherence scale; the second argument is the jet mass. 
To lowest order, 
$J_g (p^2 , k^2) = \delta (k^2) + \dots$. In this case 
in Eq.~(\ref{branchfact}) 
we reobtain the  phase space for the $ \gamma g g$ final state;  
then  the tree-level amplitude 
${\cal M}^{\rm{(tree)}}$ becomes proportional to  
the amplitude ${\cal M}_0$ of Eq.~(\ref{tree}), 
 and Eq.~(\ref{branchfact}) 
simply gives back the  result (\ref{borng}). 
In general, $J$ satisfies an evolution equation of the form 

\begin{equation}
J (p^2 , k^2) = \delta (k^2) + \int \alpha_s (p^{\prime 2})  
{\cal K} ( 
p^{\prime 2} , k^2) \otimes J  ( p^{\prime 2} , k^2) \;\;\;,  
\end{equation} 
where  expressions for the kernel ${\cal K}$ are known to leading and 
next-to-leading order\cite{cttw}. 
In the present discussion  we  limit ourselves to  
 considering a double logarithmic approximation to 
$J$. In this approximation\cite{coh}   
\begin{equation}
\label{doublog}
\int d k^2 \ J_g ( p^2 , k^2) \ \Theta (Q^2 - k^2) 
\approx 
\exp \left[ - {{\alpha_s 
} \over {2  \pi} } \ C_A \ \ln^2 \left( {p^2 \over Q^2} \right) \right] 
 \hspace*{0.2 cm} .   
\end{equation}

To determine the logarithmic 
contributions (\ref{hologs}) to the photon spectrum, 
we need to evaluate 
the branching formula (\ref{branchfact})  explicitly. 
We are interested   in  the angular-ordered, coherent  region 
\begin{equation}
\label{logregion}
k_1^2 \hspace*{0.1 cm} , \hspace*{0.1 cm} k_2^2 \ll 
(k_1+k_2)^2 \ll M^2 
 \hspace*{0.6 cm} .   
\end{equation}
The  key observation concerns the phase space 
available for the evolution of the jets $k_1$ and $k_2$ in this region.  
The boundary on the energy fraction $x_1$ of jet 1 comes from\cite{egmond}  

i) fragmentation of jet 1:  this gives  
 $x_1 \greatersim \sqrt{4 k_1^2 / M^2}$;

ii) recoil of jet 2: after using the 
angular ordering to approximate the phase space in Eq.(\ref{branchfact}), 
this gives $x_1 
\greatersim k_1^2 / [M^2 (1-z)]$.

For $z \to 1$ the tightest 
 constraint is  set by ii). 

By evaluating the phase space in Eq.~(\ref{branchfact}) explicitly 
we obtain\cite{egmond}    
\begin{eqnarray}
\label{unintG}
{1 \over \Gamma} \ 
{{d \Gamma} \over { d z} } &\simeq& 
{1 \over {16 (2 \pi)^3 (\pi^2 -9)}} \
\\
&\times& \int_0^1 d x_1 
  \int_0^1 d x_2   \ {\cal M}_0 (x_1 , x_2, z) \ \delta(z - 2 + x_1 + x_2)
\  \Theta(x_1+x_2 -1) 
\nonumber\\
&\times&   
 \int_0^\infty d k_1^2   \ J_g \left( M^2 (1-z) , k_1^2 \right) 
 \ \Theta(M^2 x_1 (1-z) -k_1^2) \ \Theta(M^2 x_1^2/4 -k_1^2)
\nonumber\\
&\times&  
 \int_0^\infty d k_2^2   \ J_g \left( M^2 (1-z) , k_2^2 \right) 
 \ \Theta(M^2 x_2 (1-z) -k_2^2) \ \Theta(M^2 x_2^2/4 -k_2^2)
\hspace*{0.6 cm} ,  
\nonumber  
\end{eqnarray}
where ${\cal M}_0$ is given in Eq.~(\ref{tree}).

Eq.~(\ref{unintG})  allows us to 
discuss  the logarithmic behaviors in the endpoint region 
by using the results for the jet mass 
distributions. 
By expanding 
the double logarithmic expression 
(\ref{doublog})  in powers of $\alpha_s$ 
we see that the photon spectrum contains corrections 
involving integrals of the form 
\begin{equation} 
\label{logx1} 
\int_{1-z}^1 d x_1 \ln \left( {{ M^2  (1-z) } \over { M^2 x_1 (1-z) }} 
\right) 
= - \int_{1-z}^1 d x_1 \ln x_1 
 \hspace*{0.6 cm} .         
\end{equation} 
That is,   logarithmic contributions in $x_1$ arise, 
which are important at the kinematic limit $x_1 \to 0$, 
but these never 
give rise to 
logarithms of $(1-z)$ in the spectrum. All $\ln(1-z)$ cancel because 
of the form  of the constraint on the jet mass, which in 
turn is a consequence of the angular ordering of the gluon radiation. 

Therefore, 
higher orders of perturbation theory do not contribute a 
Sudakov  
suppression of the photon spectrum in the endpoint region. They give 
rise to a constant shift compared to the  lowest order answer.

The picture underlying 
this result can be understood 
in simple terms. 
In the boundary kinematics 
the photon recoils against two almost-collinear gluon jets. 
 The cancellation of Sudakov corrections  reflects the fact that 
color is neutralized already at the level of this two-jet configuration.  
  This situation may be   
  contrasted with the situation one encounters in the 
  decay of an electroweak gauge  boson into jets\cite{coh,cttw}. 
Here  Sudakov corrections arise precisely from the presence of color 
charges in two-jet kinematics. 

This picture also indicates that no  cancellation should  
occur  for the color-octet Fock state in the quarkonium. In this case, 
we expect the usual Sudakov suppression to take place near the 
endpoint of the photon  spectrum. Then one of the consequences 
of the result presented here  concerns the ratio of the  
 color-octet   to the  color-singlet contributions. 
 This ratio  will be smaller than expected from the 
power counting  in $v$ and $\alpha_s$ 
obtained by truncating perturbation theory
to  fixed  order\cite{rothwise}, owing to the 
different  high order behavior of the perturbation 
series  in the two cases. 

To conclude, we observe that resummation formulas of 
the type (\ref{branchfact}),(\ref{unintG}) 
can be expanded to fixed order in $\alpha_s$ and matched 
with leading and next-to-leading\cite{krae} results to obtain 
improved  predictions,  valid over a wider range of 
photon energies.  
If these formulas are combined with   models for the 
infrared behavior of the 
strong coupling\cite{dikeshif}, they can be  used  
to model the nonperturbative shape functions\cite{shape} that parameterize 
power-like  corrections\cite{montfield,consfie} near the endpoint.  
Taking account of   effects from the soft region  will likely  
influence\cite{wolf,krae} the 
estimate of the uncertainty on the determination of 
$\alpha_s$ from quarkonia decays\cite{montfield,cleo,kobel}. 

\nonumsection{Acknowledgements}
\noindent
I thank S.\ Catani for collaboration on quarkonia decays and for 
discussion, and the organizers of the QCD session at DPF2000 for 
their invitation.

\nonumsection{References}

\end{document}